\documentclass[runningheads]{llncs}
\AtBeginDocument{%
  }
\usepackage[utf8]{inputenc}
\usepackage{algorithm}
\usepackage[noend]{algpseudocode}
\usepackage{enumitem}
\usepackage{graphicx}
\usepackage{multirow}
\usepackage{booktabs}
\usepackage{bm}
\usepackage[compatibility=false]{caption}
\usepackage{subcaption}
\usepackage{listings}
\usepackage{fullpage}

\usepackage{rotating}

\title{How Bitcoin Forms Its Network: \\ 
Peer-Table Sampling and Structural Properties} 
\titlerunning{How Bitcoin Forms Its Network}

\author{Taki E.M. Abedesselam\inst{1,2} 
\and Antonio Cruciani\inst{3}
\and Fabio Giacomelli\inst{1,2}
\and \\ Lucianna Kiffer\inst{4}
\and Francesco Pasquale\inst{1}
}
\authorrunning{T.E.M. Abedesselam et al.}
\institute{University of Rome ``Tor Vergata'', Rome, Italy \\ 
\email{abedesselam@ing.uniroma2.it, fabio.giacomelli@uniroma2.it, pasquale@mat.uniroma2.it}
\and University of Camerino, Camerino, Italy
\and Aalto University, Espoo, Finland \\
\email{antonio.cruciani@aalto.fi}
\and IMDEA Networks, Madrid, Spain \\
\email{lucianna.kiffer@networks.imdea.org}}

\date{}

\begin{document}

\maketitle

\begin{abstract}
The global structure of P2P networks underlying modern cryptocurrencies is
hidden by design: each node only knowns its neighbors and maintains a local
\textit{peer table} of IP addresses. The peer-table of a node can be seen as
the node's ``local view'' of the set of nodes currently in the network: it is
constantly updated with information collected from the neighbors and it is used
by the node to establish new connections when needed. The maintenance rules of
the nodes' peer tables determine the global structure of the network and its
evolution. Even though the global structure is unknown to any node of the
network as well as to any external observer, the rules for the exchange of
information between neighbors are compatible with the design and use of network
\textit{crawlers} that can query the nodes and extract some information about
their peer tables.

In this paper we first analyze the data we crawled from nodes of the Bitcoin
and Dogecoin P2P networks, to collect information about the distribution of IP
addresses in the peer tables and to estimate the evolution of network size and
churn rate; we then use the estimates to setup the parameters for a simulation
of the Bitcoin network and we analyze the evolution of the network structure
that we get from the simulation. Overall, our results show that Bitcoin’s
peer-table maintenance rules induce non-uniform, heavy-tailed visibility
patterns while nevertheless generating a well-connected and structurally robust
network.

\end{abstract}

\section{Introduction}
Communication in modern cryptocurrencies is based on the distributed formation
and maintenance of an unstructured P2P network. Cryptocurrencies like Bitcoin
and derivatives rely on custom network-formation processes that are the product
of several upgrades introduced during the years to face network-level attacks
or discovered vulnerabilities~\cite{heilman2015eclipse,neudecker2018network}.

At the very core of these processes there is the following basic mechanism for
peer discovery: when a node $u$ runs the software for the first time, it
bootstraps by contacting one or more DNS seed servers whose addresses are
hard-coded in the software. These DNS seeds return a list $L$ of “likely
active” nodes and record the IP address of $u$ so that $u$ may be included in
future responses to queries by other nodes;\footnote{Note that this initial
\textit{bootstrap} phase, in which a node $u$ relies on one or more DNS to
discover some of the nodes in the network, can be avoided by $u$ if somehow the
owner of the node is already aware of some other node already in the network.}
node $u$ initializes a local \textit{peer table} $M_u$ with nodes in $L$ and
attempts to establish connections to them. Every time $u$ connects to a new
node $v$, it also asks $v$ for new nodes and includes the list of nodes
sent by $v$ in its peer table $M_u$; node $u$ continues to attempt connections
from peers chosen in $M_u$ until it reaches a predetermined threshold for
\textit{outgoing connections}, while it may also accept \textit{incoming
connections} from other nodes up to a separate threshold.  The main reasons for
a node to maintain an updated peer table are (i) to enable the node to
establish new connections when its neighbor count drops below a predefined
threshold and (ii) to supply newly joining nodes with the IP addresses of
potential peers. 

It is well-known that, if a node $u$ in need of new neighbors could connect to
a node $v$ chosen uniformly at random among the nodes currently in the network,
then the resulting evolving network would be almost always well-connected
despite churn~\cite{pandurangan2003building,becchetti2023expansion}. However,
keeping up-to-date information about all nodes is costly in terms of
communication (a crawler must run continuously) and can only be done
approximately: nodes that leave do not notify others, and new nodes are not
discovered instantly. Moreover, the mechanisms for maintaining these peer
tables have to be resilient to potential attacks, hence they are often complex
and heuristically designed to hide as much information as possible while
allowing the network to be stable and robust despite churn and possible
malicious behaviors of some of the nodes. In particular, the global network
structure is unknown to any node of the network as well as to any external
observer and the peer-discovery process is designed so that the content of a
node's peer table is never fully disclosed to other nodes, but can only be
partially observed and inferred by collecting the addresses that the node
reveals.

\subsection{Our contribution}
We first provide a measurement study of peer discovery dynamics in the Bitcoin
and Dogecoin\footnote{Dogecoin is a code fork of Bitcoin that implements the
same peer-discovery protocol~\cite{dogecoincore}, making it a natural
comparison point at roughly an order of magnitude smaller scale.} networks
based on several months of \textit{crawl data}: A crawler
connects to participating nodes and interacts with them using the standard
peer-discovery protocol, i.e., the crawler requests a sample of the node's
known peer addresses and the node returns a message containing a random subset
of entries from its internal peer table, with a default maximum of $1000$
addresses. By aggregating repeated samples of nodes' peer tables we estimate
the empirical distribution of IP-address frequencies and we observe that it is
heavy-tailed. We use daily crawl observations also to estimate network size and
churn rates and our analysis shows that the daily churn rate is approximately
$5\%$ for both the larger Bitcoin network (with approximately $10K$ reachable
nodes) and the smaller Dogecoin network (with approximately $600$ nodes).

We then design and run a simulator for the evolution of the Bitcoin P2P
network. In the simulator we set the rate at which nodes join and leave the
network to match churn rates and network sizes estimated using our crawl data.
We use the simulator to study quantities that are not observable in the real
Bitcoin P2P network, in particular we focus on the analysis of the internal
peer-table dynamics of each node and on the evolving global network structure.
In the simulation we find that the distribution of IP-address frequencies in
the peer tables of the nodes is heavy-tailed, as observed in the crawl data:
since in our simulation every node is running the same software with the same
default parameters, this highlights that the heavy-tailed distributions of
peer-table entries observed in Bitcoin and Dogecoin are a direct consequence of
the peer selection and maintenance mechanisms in the protocol and cannot be
attributed to, e.g., the heterogeneity in node activity. Despite this skew, the
evolving network resulting from the simulation remains highly connected and
robust, demonstrating that protocol-level design choices alone can generate
both biased visibility and resilient network structure. In fact, it turns out
that the evolving network resulting from the simulation exhibits a very small
diameter (around $5$) and a spectral gap of approximately $0.5$, indicating
very good connectivity properties.

Beyond the present analysis, the simulation framework is readily extensible to study protocol modifications, adversarial scenarios, or further heterogeneous node deployments.

\subsection{Related work}
In the theoretical literature, the ``peer sampling'' problem in P2P network is
often referred to as ``membership problem'' and has been studied, for example,
in~\cite{allavena2005correctness,jelasity2007gossip,bortnikov2008brahms,guerraoui2024peerswap,bakhshi2009analytical,jelasity2003newscast}.
In~\cite{allavena2005correctness} the authors present a new algorithm for local
view maintenance and they give bounds on the time it takes to have a network
partition: in particular, they show that the expected time to partition is at
least the square of the view size. In~\cite{jelasity2007gossip} the authors
introduce a generic framework for peer sampling and empirically study the
properties of local views. In~\cite{bortnikov2008brahms} the authors give a new
membership algorithm that tolerates Byzantine attacks. Their algorithm requires
the local view of each node to be $\Omega\!\left(n^{1/3}\right)$; they prove
that the local view of each node converges to a uniform sample of the
population, but they do not analyze the correlation between the local views of
different nodes. More recently, \cite{guerraoui2024peerswap} proposes a new
peer-sampling protocol and proves that, for each node, the distribution of the
peer returned by the sampler converges to uniform, with bounds on the
convergence time; again, the dependence between samples obtained at different
nodes is not characterized.  Analytical models of gossip-based membership and
view-exchange dynamics are also investigated
in~\cite{bakhshi2009analytical,jelasity2003newscast}.

On the measurement side, several works have studied the P2P overlays of
Bitcoin and other cryptocurrencies. In~\cite{deshpande2018btcmap} the authors
develop \emph{BTCmap}, a crawler that collects local address databases from
reachable peers and, together with emulated connections, reconstructs
near-real-time snapshots of the inferred Bitcoin topology. In
\cite{li2022bitcoin} the authors build a measurement system that discovers
millions of advertised addresses and identifies a few thousand reachable nodes,
reporting on properties such as service types, client versions, geographic
distribution, and stability of reachable peers. In~\cite{essaid2023nodeprobe}
the authors introduce Node--Probe, a recursive scanning technique used to
collect long-running snapshots of the Bitcoin network and to infer structural
properties such as community structure and the role of long-lived nodes. In
\cite{kiffer2025multicoins} the authors present a cross-network study of 36
public blockchain systems, measuring, among other things, the number of
reachable peers, churn, and connectivity characteristics across different
discovery mechanisms. All these measurement studies treat peer tables and the
address manager as black boxes: they use reachable peers and advertised
addresses as an oracle for (inferred) topology, but they do not explicitly
analyze the sampling properties or biases of the underlying peer-table
maintenance algorithms.

On unreachable peers and NATed nodes, several works have analyzed what can be
inferred from address announcements in the Bitcoin network.
In~\cite{grundmann2022short} the authors study peer announcement messages
(\texttt{ADDR}) and propose the PAL (Passive Announcement Listening) method to
estimate the total number of active peers and the fraction that are
unreachable; their results indicate that the majority (approximately $75\%$) of
nodes are unreachable (e.g., due to NATs or firewalls) and yet still contribute
to transaction and block propagation. In~\cite{saad2021synchronization} the
authors perform a root-cause analysis of deteriorating Bitcoin network
synchronization and, among other factors, quantify the size of the unreachable
population relative to reachable peers and show that the addressing protocol
does not distinguish between reachable and unreachable addresses, leading to
many failed outgoing connection attempts and a reduced effective out-degree for
honest nodes.

On the security side, several works have studied attacks that exploit Bitcoin's
peer-discovery and address-management mechanisms. In~\cite{heilman2015eclipse}
the authors describe eclipse attacks on the Bitcoin P2P network, showing how
details of Bitcoin Core's address manager (e.g., new/tried buckets and eviction
rules) can be used to fill a victim's peer table with attacker-controlled
addresses and to monopolize its connections. In~\cite{tran2020erebus} the
authors present the EREBUS attack, a stealthy partitioning attack in which an
AS-level adversary gradually influences peering decisions and becomes a
man-in-the-middle for all of a victim's connections, without relying on overt
BGP hijacks. In~\cite{yang2022bhe} the authors propose the BHE attack, which
combines BGP hijacking with control over TCP handshakes to bias which peers are
accepted into the victim's new and tried tables and to maintain long-term
occupation of its connection slots. These works all highlight that Bitcoin's
membership mechanism is non-uniform and history-dependent enough to be
exploitable, but they focus on adversarial scenarios and do not characterize
the unbiased, steady-state distribution of peers in honest nodes' tables.

\subsection{Organization of the paper}
In Section~\ref{sec:background} we synthesize the operating mechanisms related
to the network formation process of a Bitcoin node running the
Bitcoin-core~\cite{bicoincore-github} software under normal conditions, i.e.,
without applying any modifications to the source code nor any specific custom
configuration. The details of such mechanisms provide some background for the
results in Sections~\ref{sec:empirical} and~\ref{sec:simulations}.  In
Section~\ref{sec:empirical} we analyze the data we collected for two
cryptocurrency P2P network: \textit{Bitcoin}~\cite{nakamoto2008bitcoin} and,
for comparison, one of its several forks with a large market capitalization,
\textit{Dogecoin}.  In Section~\ref{sec:simulations} we describe our simulator
based on Bitcoin-core and we analyze the results of the simulations.

\section{Bitcoin-core network formation primer}\label{sec:background}
In Bitcoin-core~\cite{bicoincore-github}, the main implementation of Bitcoin
and the one we use as a reference for our simulator in
Section~\ref{sec:simulations}, addresses collected during bootstrapping or from
peers (see Subsections~\ref{ssec:bootstrap} and~\ref{ssec:peerdiscovery}) are
handled by the address manager, \texttt{AddrMan}, and stored in two tables:
table \texttt{vvTried} for addresses the node has itself connected to
successfully, that consists of $256$ buckets of $64$ entries each, and table
\texttt{vvNew} for all other addresses, that consists of $1024$ buckets of $64$
entries each. The assignment of an address to a entry is done with a
cryptographic hash function salted with a private random $256$-bit key, so that
the position of an address in the table of a node is unpredictable for an
attacker.

\subsection{Bootstrap}\label{ssec:bootstrap}
The first time a node executes Bitcoin-core, the node's \texttt{vvNew} and
\texttt{vvTried} tables are empty, so \texttt{AddrMan} first queries a set of
\textit{DNS seeds} ($8$ hostnames on mainnet), and it inserts in \texttt{vvNew}
at most $32$ addresses from each seed, so that no single seed dominates the
table. If the table is still empty after the queries to the DNS seeds, the node
falls back to a hardcoded list of \textit{fixed seed nodes}.  Once the
\texttt{vvNew} table contains some addresses, the node starts trying to
establish connections with them.

Notice that an operator who already knows some ``trusted'' Bitcoin node can
bypass the bootstrapping phase: for example, with the option \texttt{-connect}
the operator can pass one or more addresses as parameters and the node will
exchange information only with the given addresses for its entire lifetime,
disabling peer discovery entirely, while with the option \texttt{-addnode}  the
operator can add one or more persistent addresses without disabling DNS
queries and peer discovery.

\subsection{New connections}
In the default configuration, a node has a maximum of $125$ connections. Ten
connections are established by the node itself and they are called
\textit{outbound connections}, more precisely a node establishes eight
\texttt{OUTBOUND\_FULL\_RELAY} connections, that are used to exchange blocks,
addresses and transactions, and two \texttt{BLOCK\_RELAY} connections, that are
used to exchange only blocks.

Candidate addresses for outbound connections are drawn by \texttt{AddrMan} from
tables \texttt{vvNew} or \texttt{vvTried} according to the following procedure:
when both tables are non-empty the node chooses \texttt{vvNew} or
\texttt{vvTried} with equal probability, then it picks a bucket uniformly at
random from that table and a random position within the bucket, then the node
takes the address in the first non-empty slot at or after that position,
wrapping around if needed (if the bucket is empty the sampling is repeated).
Bitcoin-core tries to spread nodes' outbound connections across different
\textit{network groups} to avoid that two or more outbound connections are
likely controlled by the same entity or located on the same network; in recent
versions of Bitcoin-core, two addresses are in the same network group if they
belong to the same Autonomous System (AS)~\cite{hawkinson1996guidelines}. A
\textit{diversity filter} is thus applied to the selected candidate address
that rejects any address in the same network group of one of the node's current
outbound peers. The selection that produces this candidate is itself
probabilistic: a table is chosen by a fair coin flip, a uniformly random bucket
and position are drawn, and the entry found there is returned with a
probability that falls by a factor $0.66$ per failed connection attempt
recorded since the addresses last successful handshake, up to eight such
failures beyond which the penalty saturates, and by a further factor $100$ if is
the address was attempted within the last ten minutes. This probability is not
stored with the address: its recomputed at each draw from the attempt counter and the last-try timestamp. If the entry is not returned, a \emph{new} candidate is drawn and its acceptance probability is scaled up by a factor
$1.2$, a factor local to the invocation that serves only to bound the loop.
Once an address is returned and passes to the diversity filter, the node initiates the connection handshake and tries to establish the connection with the peer. During the handshake, if the address comes from the \texttt{vvNew} table the node promotes it to the \texttt{vvTried} table. In case the designated position in the \texttt{vvTried} table is already occupied, a \textit{collision handling} procedure is initiated that decides whether to keep the previous address or to replace it with the new address.

\subsection{Peer discovery}\label{ssec:peerdiscovery}

Bitcoin nodes use two ways to share addresses between neighbors:
\textit{\texttt{GETADDR} requests} (\textit{pull}-type information
dissemination) and \textit{self-advertisement} (\textit{push}-type).

Every time a node establishes a new outgoing connection, after the handshake it
sends a \texttt{GETADDR} request. Each node prepares the response to
\texttt{GETADDR} requests, called \texttt{cache}, every $24$ hours on average
by picking addresses uniformly at random from the union of all addresses in the
two tables, \texttt{vvTried} and \texttt{vvNew} (an extra ``quality'' filter is
applied that prevents to select addresses that are marked \textit{terrible}).
The size of the \texttt{cache} is the minimum between $1000$ and $23\%$ of its
known IP addresses. Each \texttt{cache} is kept for a random lifetime between
$21$ and $27$ hours and served unchanged to every requester during that period,
so repeated queries reveal nothing new.  A node answers a \texttt{GETADDR}
request only when it comes from an \textit{inbound} connection, and
only once per connection. Requests on connections the node itself initiated are
ignored, which prevents an attacker from seeding a victim's \texttt{AddrMan}
and reading it back~\cite{biryukov2014deanonymisation}.

In addition, a node also periodically shares addresses with its neighbors
through self-advertisement by sending unsolicited \texttt{ADDR} messages. On
average every 30 seconds, each node sends an \texttt{ADDR} message to each of
its neighbors, containing addresses from a buffer called \texttt
{m\_addrs\_to\_send} (limited to $1000$ addresses, distinct for each neighbor
and emptied after each send). The buffer is filled from three a single bulk response sources: addresses
relayed from unsolicited \texttt{ADDR} messages received from \emph{other}
neighbors, following the rules below; a single bulk response to a \texttt
{GETADDR} request, which clears the buffer before refilling it; and the node's
own address, included on average every $24$ hours. When the buffer is already
full, a new address replaces a uniformly random existing one rather than being
dropped. To control the rate at which addresses received from a neighbor are
processed, each node maintains an associated \texttt{m\_addr\_token\_bucket}
counter. This counter starts at 1 and increases by 0.1 per second since it was
last updated, capped at 1000 (exceeding 1000 only when the node sends a \texttt
{GETADDR} request). When the node receives an \texttt{ADDR} message from a
neighbor, its addresses are shuffled and processed one at a time, each
consuming one token; addresses arriving with no tokens available are discarded.
For example, if a message contains 10 addresses but only 9 tokens are
available, a uniformly random subset of 9 is processed. Each processed address
that lies in a network the node can reach is then handed to the address
manager: if it is not already known, a new entry is created in the \texttt
{vvNew} table with a two-hour timestamp penalty; if it is known, its last-seen
timestamp is refreshed and its services merged, with no further treatment for
entries already in \texttt {vvTried}. Additionally, addresses are only relayed
(appended to the \texttt {m\_addrs\_to\_send} buffer of up to two \emph{other}
neighbors) if its timestamp is less than $10$ minutes old, the node has not
sent \texttt{GETADDR} to that peer, the message carried at most $10$ addresses,
and the address is routable. The last two conditions confine relay to small
unsolicited announcements rather than bulk table dumps. Reachable addresses go
to $2$ neighbors and unreachable ones to $1$ or $2$, chosen as the highest
ranking peers under a keyed hash of the address, the peer identifier and a $24$
hour epoch whose phase is itself derived from the address, so one address
travels the same way for a whole day. Together with a per-neighbour filter of
addresses already sent, this prevents a peer from earning extra propagation by
announcing itself repeatedly.
\section{Empirical analysis}\label{sec:empirical}

We collect daily crawl data from the Bitcoin and Dogecoin peer-to-peer networks
over two three-month periods (July--September 2025 and February--April 2026); full details of the crawling
methodology are given in Appendix~\ref{app:data_collection}. We use this data
in two stages: first, to estimate network size and churn rates that inform the
parameters of our simulations (Section~\ref{sec:simulations}); second, to
characterize peer-table properties of the real networks as an empirical baseline
for comparison with our simulation results.

\subsection{Network size and churn}

We use daily crawl data to summarize network size and node availability dynamics. These measurements provide empirical inputs for the churn model used in our simulations, in which nodes independently join and leave the network with fixed probabilities.

\subsubsection{Daily network size.}
For each day, we record whether a node responded to the crawler
(\emph{crawled}) and whether it responded to at least one in-protocol
\texttt{PING} message (\emph{active}). Bitcoin is substantially larger, with a
median of approximately 10{,}000--10{,}700 active nodes per day across both
measurement periods, while Dogecoin is roughly an order of magnitude smaller at
approximately 590--650 active nodes per day. These counts are consistent across
both periods approximately six months apart, indicating that both networks are
in a stable regime. Full quantitative details, including total unique addresses
and IPv6 fractions, are given in Table~\ref{tab:network-summary} and
Figure~\ref{fig:daily-discovered-active} in Appendix~\ref{app:session_length}.

\subsubsection{Node availability and churn.}
\begin{figure}[h]
\centering
\includegraphics[width=.75\columnwidth]{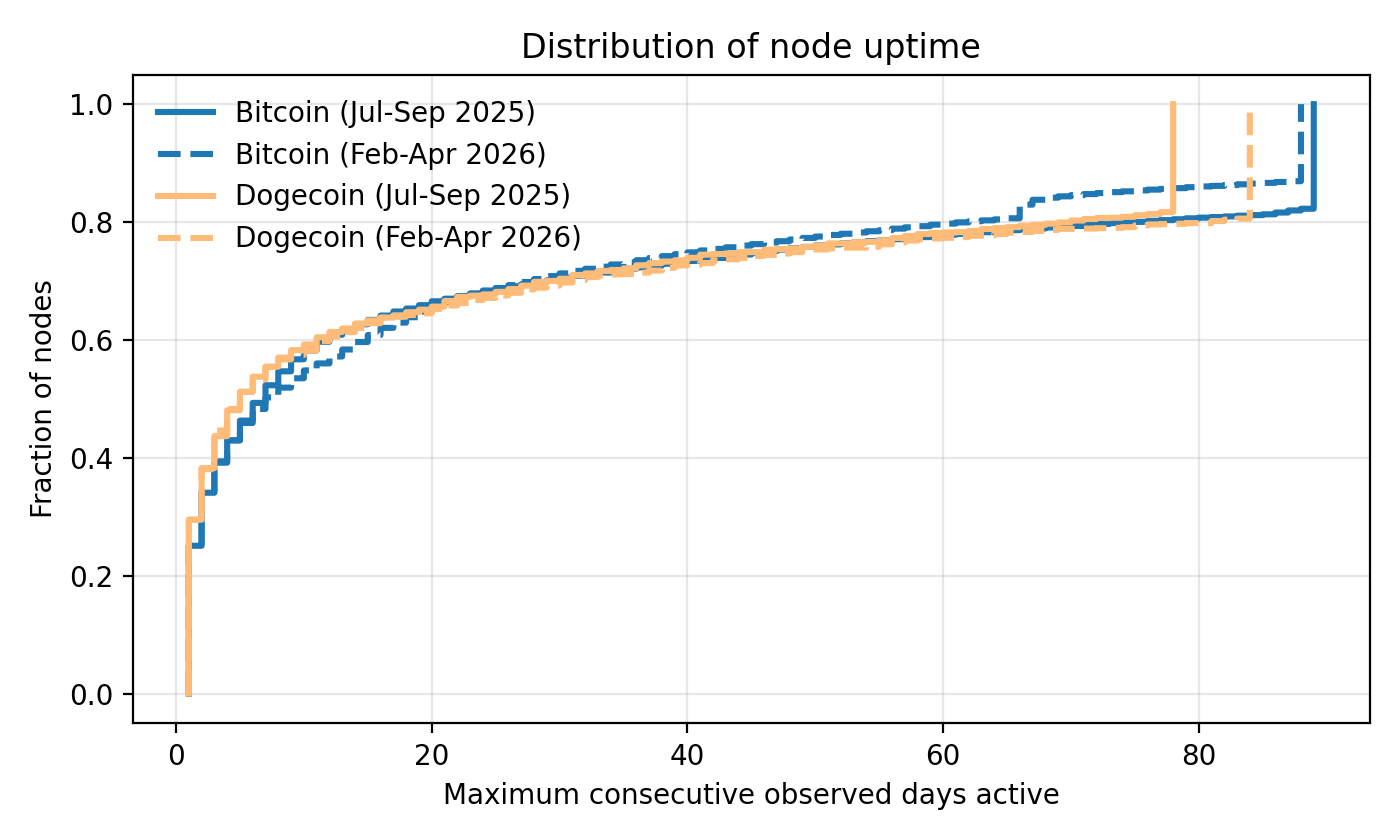}
\caption{Cumulative distribution of node uptime measured as total days active during the measurement period. Approximately 20\% of nodes in each network remain active for the entire observation window, with the remainder exhibiting varying degrees of intermittent participation.}
\label{fig:uptime}
\end{figure}

We begin by measuring per-node uptime as the total number of days each node is observed active over the measurement period. Figure~\ref{fig:uptime} shows the resulting distribution: approximately 20\% of nodes remain active throughout the observation window, while the majority exhibit intermittent participation.

To better understand these intermittent participation patterns, we analyze the structure of node sessions. We define a session as a continuous sequence of active days, Figure~\ref{fig:session-dynamics} (Appendix~\ref{app:session_length}) shows that most nodes appear in a single session and that inter-session gaps are heavily concentrated at exactly one day, suggesting that most brief absences likely reflect measurement artifacts or transient network interruptions rather than genuine node departures and returns. We therefore adopt a grace-period definition: gaps of up to one day are treated as within-session, and churn is the fraction of nodes active on day~$t$ that are absent for at least two consecutive days (day~$t{+}1$ and day~$t{+}2$).

\begin{figure}[h]
\centering
\includegraphics[width=.8\columnwidth]{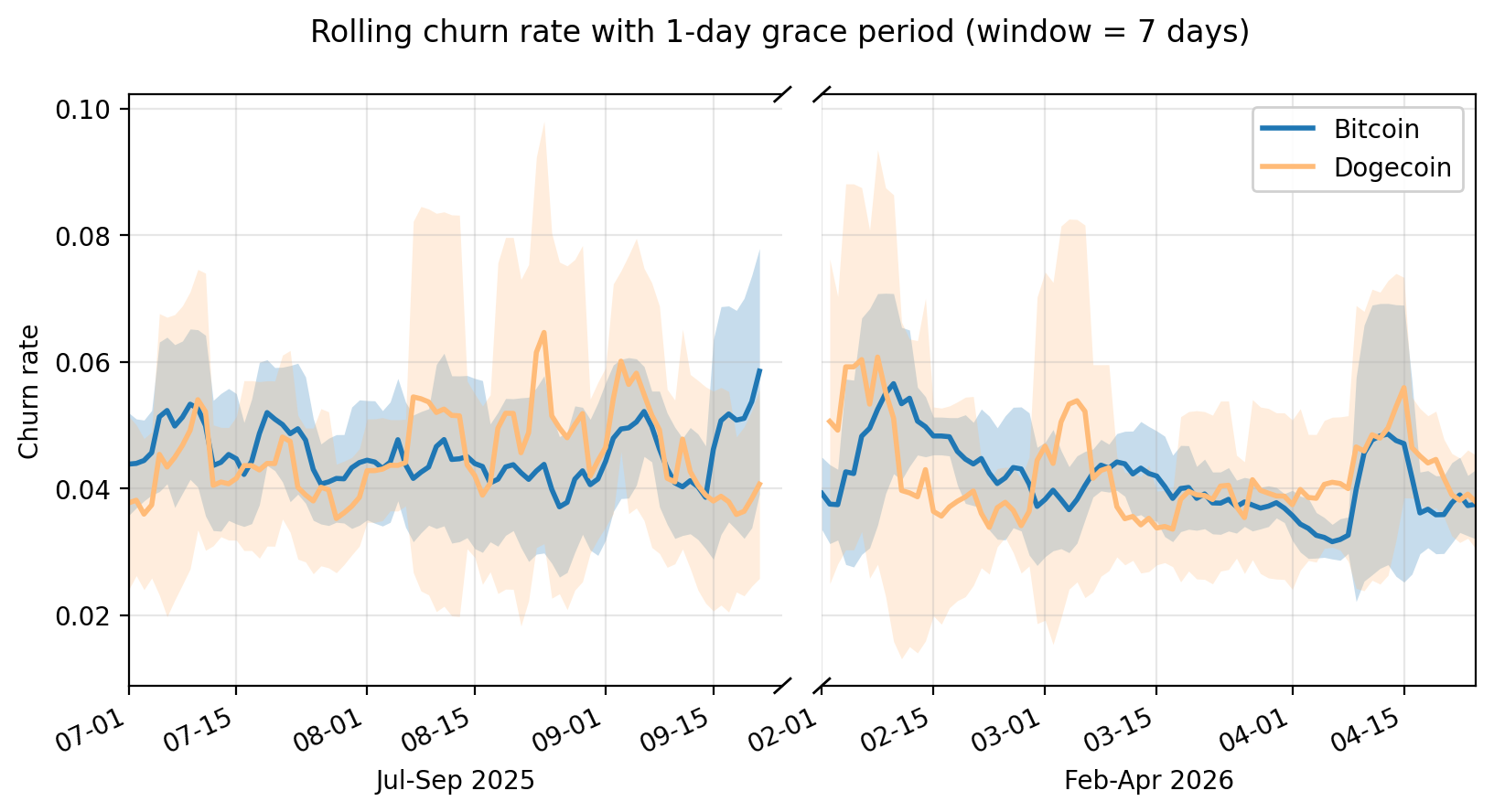}
\caption{Rolling 7-day mean churn rate for Bitcoin and Dogecoin, computed with a one-day grace period. Shaded regions indicate one standard deviation above and below the mean. Churn is defined as the fraction of nodes active on day~$t$ that remain absent for at least two consecutive days.}
\label{fig:daily-churn-rolling}
\end{figure}

Figure~\ref{fig:daily-churn-rolling} shows the rolling 7-day mean churn rate for both networks. While both networks exhibit short-term variability, churn remains concentrated around 0.05 across both measurement periods, and we use this value to parameterize our simulations.

\subsection{Peer table sampling and address visibility}

We next turn to measurements derived from nodes' advertised peer tables. Since peer tables are never fully disclosed, our analysis relies on repeated partial observations: each \texttt{GETADDR} response returns a uniformly random subset of up to 1{,}000 addresses from a node's internal peer table. We treat each such response as an independent sample. After filtering to responses containing at least 800 addresses to exclude nodes with non-IPv4/IPv6 configurations (see Appendix~\ref{app:data_cleaning}), we pool all responses across all days within each measurement period and count, for each address, the number of responses in which it appears. This gives an empirical address-frequency distribution reflecting how visible each address is across the network's peer tables.

\begin{figure}[h]
  \centering
  \includegraphics[width=\linewidth]{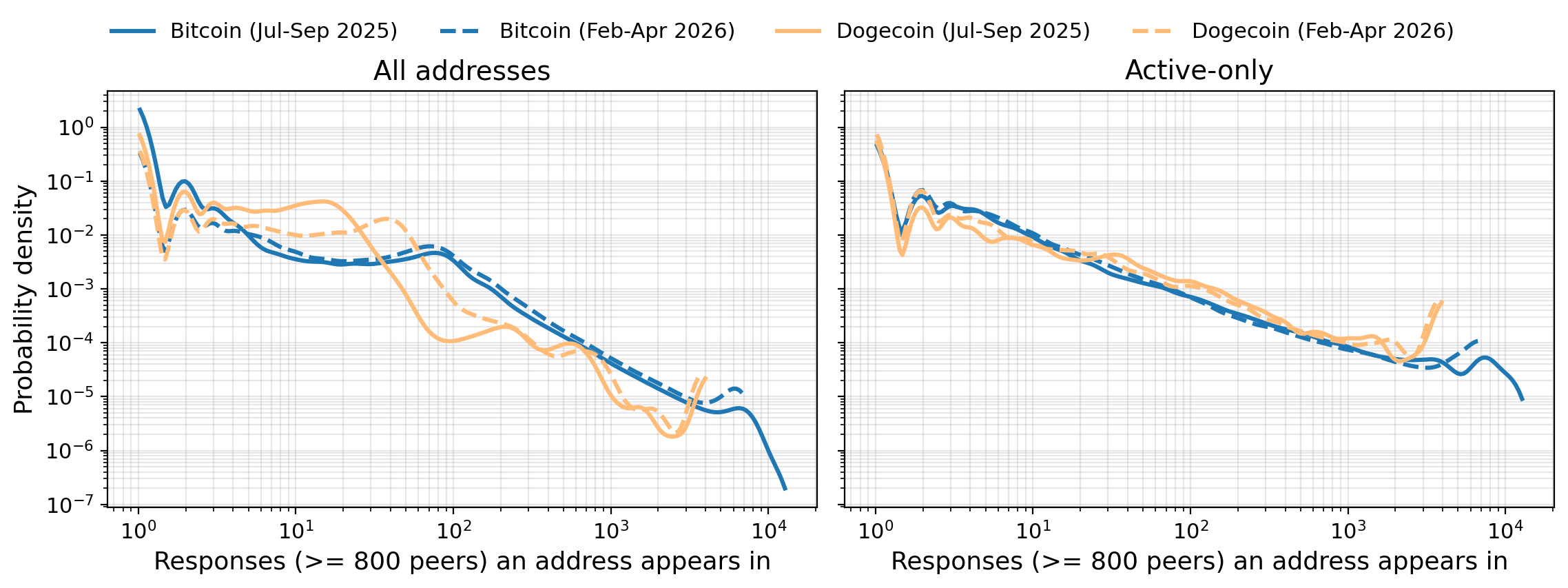}
  \caption{CCDF of address frequencies across pooled peer-table samples for Bitcoin and Dogecoin, across both measurement periods. Left: all addresses. Right: restricted to addresses observed active during the measurement period. Active addresses are systematically more visible, appearing in a greater fraction of samples than the overall population.}
  \label{fig:address-frequency}
\end{figure}

This behavior is consistent across both measurement periods and both networks, and is the empirical property we compare against our simulation results in Section~\ref{sec:simulations}.
\section{Simulations}\label{sec:simulations}
In this section, we present our network simulator designed to replicate
Bitcoin's P2P network-formation process when all nodes run Bitcoin Core with
the default settings, and we present the results obtained with the simulation.
In particular, we characterize how peer address information is distributed
across node caches and address tables, and we examine the structural properties
of the emerging network topology.

\subsection{Network Model}\label{sec:sim-model}
The simulated network is a dynamic graph $G(t) = (V(t), E(t))$ in which
vertices are nodes running Bitcoin Core with the default settings and edges are
established TCP connections. Nodes arrive according to a Poisson process of
rate $\lambda$; every time a node $u$ joins the network a \textit{session
length}, i.e., the length of the interval of time that node $u$ remains active
in the network, is chosen randomly. Joining rate $\lambda$ and session lengths
are chosen so that the expected churn rate in the simulation matches the one
derived in Section~\ref{sec:empirical} from the crawl data. 

In the real Bitcoin network, in addition to new nodes that join the network and
nodes that permanently leave it, there are nodes that temporarily disconnect
and then reconnect to the network with the same identity; in our simulation we
have a tunable parameter $q$ to govern that behavior: when a node leaves the
network, with probability $q$ the node is scheduled to rejoin the network in
the future, keeping its identity and its address tables.

To model the \textit{network grouping} strategy of the addresses in the real
network, in the simulation we partition the $32$-bit address space into
$n_{\mathrm{AS}}$ ``virtual Autonomous Systems'' whose sizes follow a power
law, and we use this membership wherever Bitcoin-core uses the network group of
a peer. To model reachable and unreachable nodes, in the simulation a joining
node is set reachable with probability $p_{\mathrm{r}}$ and it is otherwise
behind a network address translator; the probability $p_{\mathrm{r}}$ is chosen
so that the expected proportion between reachable and unreachable nodes matches
the proportion estimated in~\cite{grundmann2022short} for the real Bitcoin
network. Unreachable nodes refuse inbound connections, yet they still open
outbound ones, advertise their own address and relay the addresses of others.
Their addresses therefore circulate and occupy table entries even though no
connection to them can ever succeed. 

We model only \texttt{OUTBOUND\_FULL\_RELAY} connections, since block relay
connections carry no address traffic and therefore do not affect peer
discovery, which leaves $117$ inbound slots within the default limit of $125$.

\subsection{Simulation Results}
From the crawl data in Section~\ref{sec:empirical} three interesting aspects of
the peer discovery process emerge that measurement alone cannot shed light on,
since a crawler observes only what nodes choose to reveal: (i) the nature of
the addresses that populate peer tables, (ii) the origin of the gap between
discovered addresses and active nodes, and (iii) the global topology itself,
which is unknown to every participant and to every external observer. In the
simulation we can instead directly observe every address, connection and table
entry. The results presented here cover the first $120$ simulated days, during
which the network grows up to $33{,}598$ nodes, where $5{,}660$ of them are
reachable.

We measure the simulated network through three instruments. Every node rebuilds
its \texttt{GETADDR} advertisement cache daily, exactly as Bitcoin-core does,
and we record its full contents, so the union over one day defines the daily
discovered set and the union over the window the total discovered set. Every
time a node leaves the network we record its complete address manager, giving
table level statistics no crawler can obtain, since a real \texttt{GETADDR}
response exposes at most $1{,}000$ addresses of a table that we find to hold
about $51{,}000$. Finally a virtual crawler samples the caches of reachable
nodes and aggregates address frequencies per crawl (definition~A) and per node
(definition~B), replicating the pipeline of Section~\ref{sec:empirical} by
pooling all responses across the whole window, $575{,}609$ full size samples
over days $28$ to $120$, a volume comparable to the $508{,}039$ Bitcoin
responses retained for one measurement period. An address counts as active if
it belongs to a node reachable during the window, the simulated analogue of
responding to \texttt{PING}.

\subsubsection{The peer tables' content.}
In Section~\ref{sec:empirical} we observed that most addresses in sampled peer
tables were never active during the measurement window and we hypothesized that
they belong to online but unreachable nodes. The simulation confirms the
mechanism and refines the interpretation. Tables are large and mostly full,
holding on average $49{,}739$ entries in the new table against a capacity of
$65{,}536$, while the tried table holds only $1{,}215$ of $16{,}384$ slots.
Only $2.4\%$ of the entries a node holds were ever connected to successfully,
and just $17.2\%$ correspond to a node alive at the moment of measurement, with
a median of $12.4\%$. The dominant reason is not that the remaining entries
belong to currently online unreachable nodes but that they are expired
addresses of unreachable nodes. Address rotation mints about $9{,}200$ new
addresses per day, with $5.2\%$ of rotations crossing Autonomous System
boundaries, and every rotation turns a circulating address into a permanent
ghost while the node lives on under a new one. The never active entries of the
crawl data are thus mostly traces of NATed nodes' past addresses rather than
their present ones.  Tables are cleaned quickly, with cached addresses
averaging only $9.6$ days of age and $44\%$ under one week, because the address
horizon and slot displacement evict old entries steadily. Yet tables remain
full of unusable entries, because new dead addresses arrive faster than cleanup
removes them.  Freshness and usability are different axes, and table entries
are recent and useless at the same time.

\subsubsection{Why discovery sees many more addresses than nodes.}
\begin{figure}[h]
    \centering
    \includegraphics[width=0.8\columnwidth]{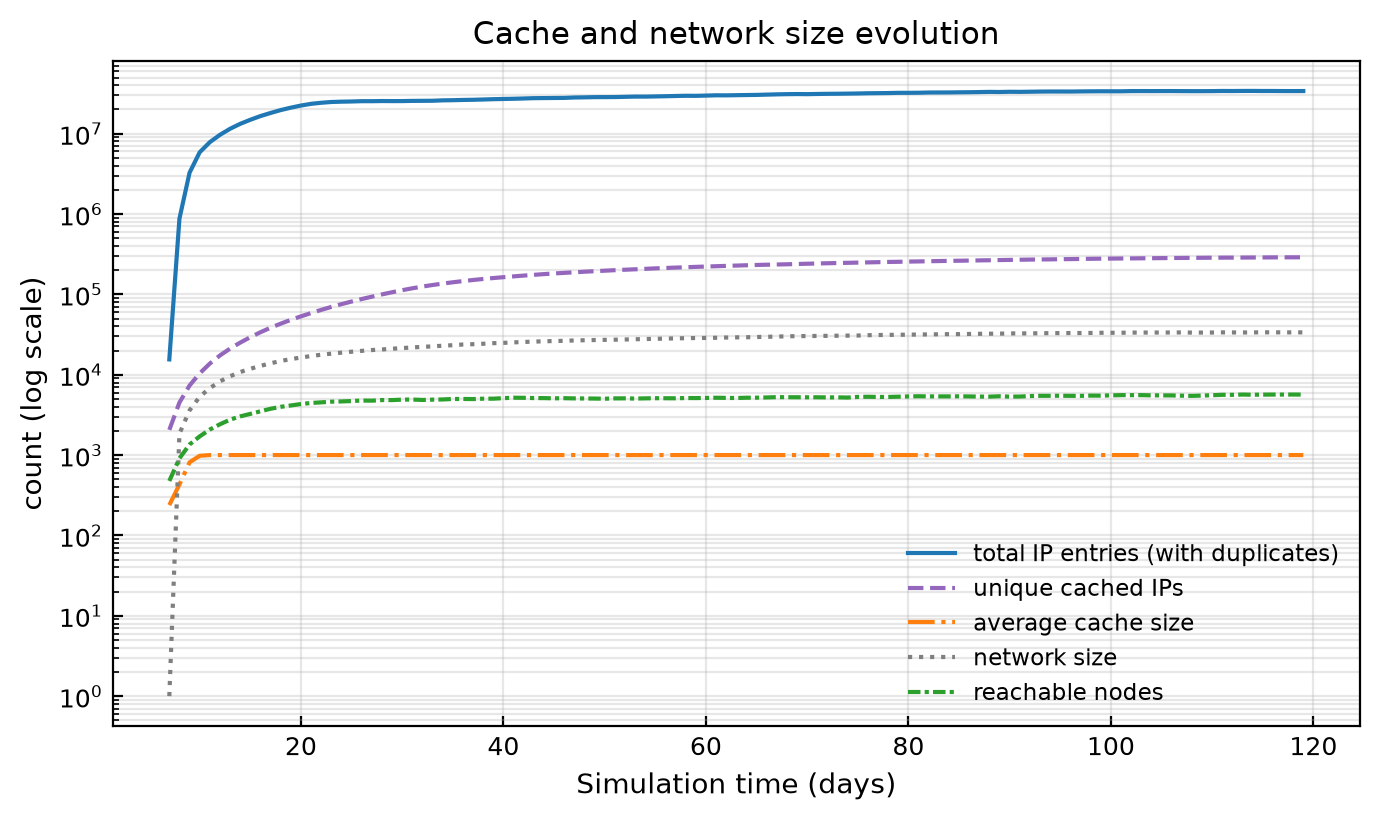}
    \caption{Cache and network size evolution over days $7$ to $119$ of the
    mixed run, on a logarithmic vertical axis. From top, total address entries
    including duplicates, distinct cached addresses, live network size,
    reachable nodes, and average cache size per response. The reachable
    population plateaus near $5{,}700$ while the set of circulating addresses
    keeps growing.}
    \label{fig:cache_growth_network_size}
\end{figure}

\begin{figure}[t]
    \centering
    \includegraphics[width=0.8\columnwidth]{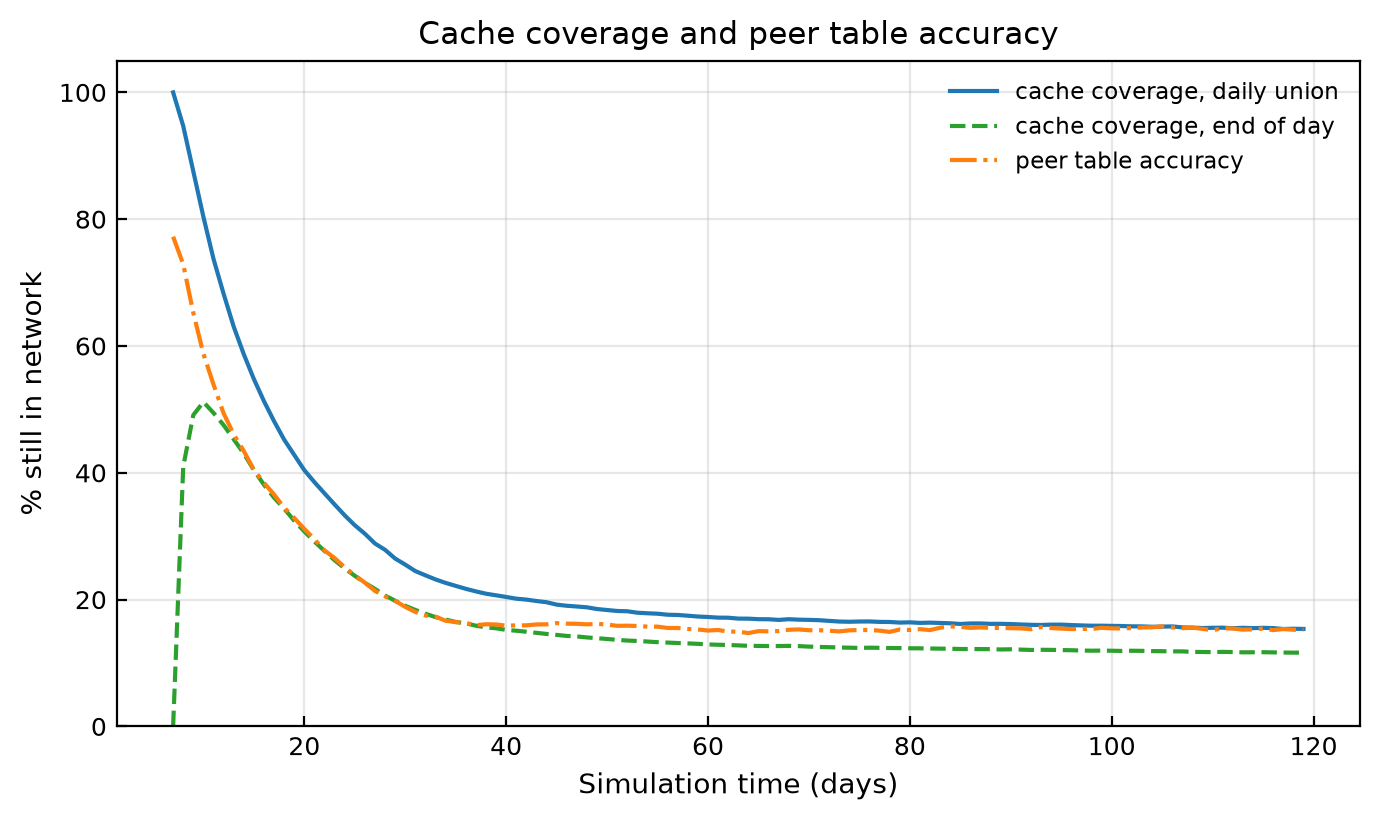}
    \caption{Fraction of cached addresses and of peer table entries that belong
    to a live node, per simulated day. Cache coverage is measured against the
    union of nodes alive at any point during the day (solid) and against the
    end of day snapshot (dashed), peer table accuracy against the end of day
    snapshot (dash dot).}
    \label{fig:network_coverage_and_accuracy}
\end{figure}

The crawl data reports $275{,}344$ discovered addresses per day against
$10{,}297$ active nodes (Table~\ref{tab:network-summary}), a ratio of about
$27$. The simulation reproduces this regime and exposes its cause. By day $119$
the daily discovered set contains $289{,}377$ distinct addresses against a live
population of $33{,}655$, a ratio of $8.6$ against all nodes and $51$ against
the $5{,}660$ reachable ones, and the total discovered set reaches $913{,}079$
distinct addresses over the window (Figure~\ref{fig:cache_growth_network_size}).
Correspondingly, the share of cached addresses belonging to any live node
decays from $45.3\%$ at day $13$ to $11.6\%$ at day $119$
(Figure~\ref{fig:network_coverage_and_accuracy}). The decay is the equilibrium
consequence of a bounded population of stable reachable addresses being diluted
by an unbounded stream of rotating unreachable ones.

The virtual crawler permits a shape level comparison with the visibility
measurements of Section~\ref{sec:empirical}, though not a direct superposition,
since the two differ in scale, in age and in how peers are identified. We
therefore compare only dimensionless features of the pooled response
distributions (see Figure~\ref{fig:crawl_visibility} in
Appendix~\ref{sec:apx:simulation}). The central observation of
Figure~\ref{fig:address-frequency} is reproduced, in that active addresses are
systematically more visible than the population, by a factor of $4.9$ at the
median. The clear divergence is at the opposite end. In the crawl data $24.8\%$
of addresses appear in at most two responses, against $0.12\%$ in the
simulation, so the measured network carries a large reservoir of barely visible
addresses that the model does not generate. Part of that reservoir may require
\textit{heterogeneity} of nodes that our model omits (in our simulation all
nodes run with the default Bitcoin-core parameters), and part may simply be the
residue of an address history far longer than the simulated window, which a
longer run could identify. Protocol telemetry quantifies the cost of operating
in this regime (see Figure~\ref{fig:telemetry} in
Appendix~\ref{sec:apx:simulation}). All steady state table growth arrives
through unsolicited \texttt{ADDR} gossip, which is exactly the path the token
bucket governs, and by day $119$ nodes drop $62{,}065$ such entries per day, up
from $9{,}170$ at day $13$. The same growth of dead addresses appears in the
\textit{feeler} mechanism, whose purpose is to verify stored addresses. Feeler
success falls from $15.6\%$ at day $13$ to $5.4\%$ at day $119$, by which point
nodes collectively make over $23$ million verification attempts per day and
serve $97{,}878$ \texttt{GETADDR} responses.

Address visibility separates sharply by class (see
Figure~\ref{fig:ip_frequency} in Appendix~\ref{sec:apx:simulation}).  For each
day we count the number of caches holding each address, and average the
resulting distribution over the stabilized window. Across all $207{,}381$
addresses present on a typical day the distribution peaks near $40$ caches and
decays from there, with a mean of $175$. Across the $11{,}946$ that belong to
reachable nodes, $5.8\%$ of the total, it peaks instead near $500$, with a mean
of $379$. A reachable address is therefore held in about $2.2$ times as many
caches as the average one, because only a reachable address can be verified by
a connection and promoted to the \texttt{vvTried} table. Below the crossing
point at $250$ caches the distribution is almost entirely made of unreachable
addresses.

\subsubsection{The network structure.}
\begin{figure}[h] \centering
\includegraphics[width=0.8\columnwidth]{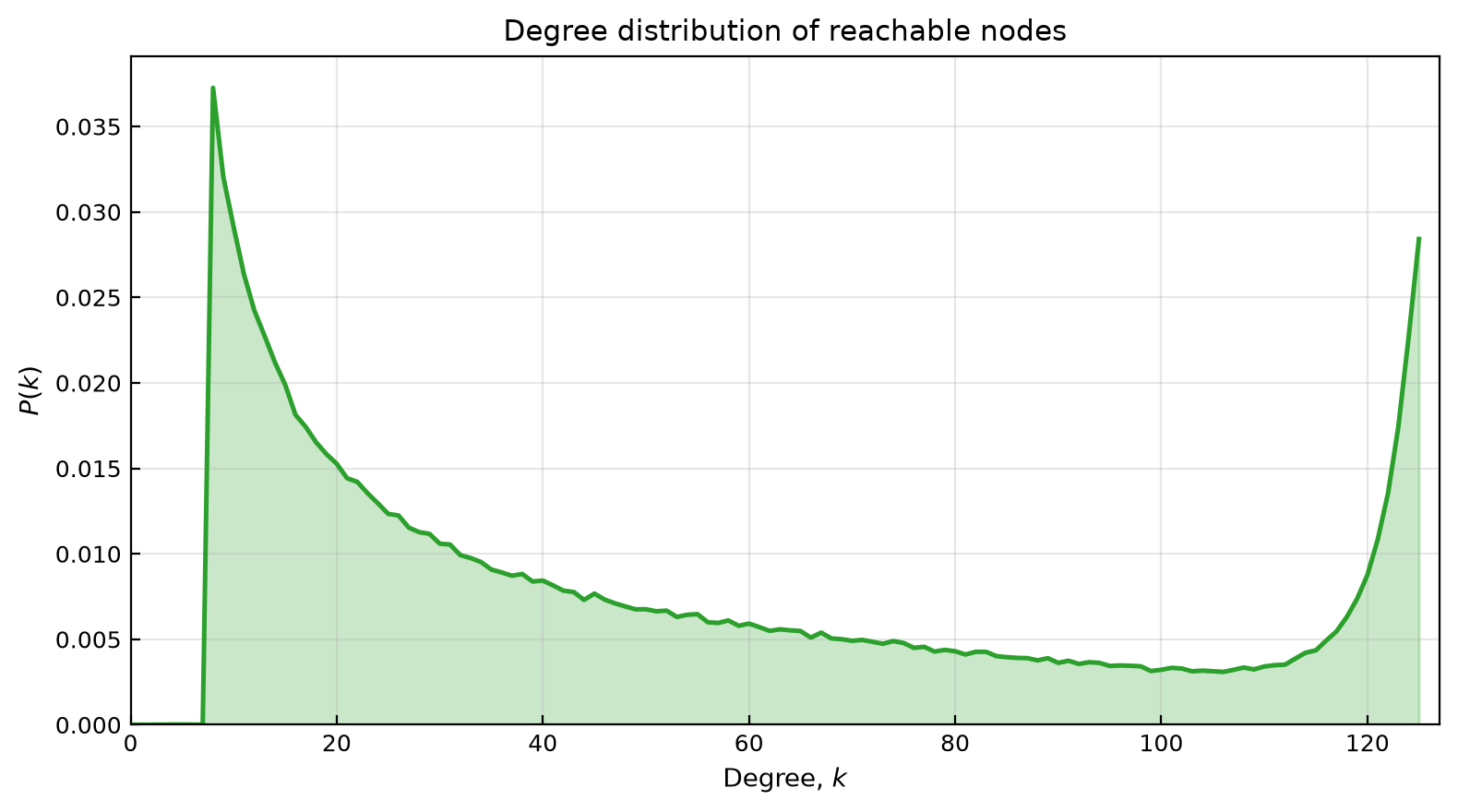}
\caption{Degree distribution of reachable nodes, normalised over the reachable
population and averaged over days $13$ to $119$. Unreachable nodes are omitted,
since they sit at degree $8$ with probability $0.9997$. Reachable nodes are
bimodal, with a second mode at the connection limit of $125$.}
\label{fig:degree_distribution} 
\end{figure}

The average degree measured in the simulation is exactly $16$ as expected,
since every node maintains $c_{\mathrm{out}} = 8$ outbound connections. What
the protocol shapes is the distribution of those connections, and it splits
sharply by class: unreachable nodes sit at degree exactly $8$ almost without
exception, $99.97\%$ of them averaged over days $13$ to $119$, because they
accept no inbound connections.  Reachable nodes span the range from $8$ up to
the connection limit of $125$, with median $39$ and mean $51.7$, and their
distribution is bimodal (Figure~\ref{fig:degree_distribution}). 

Two unreachable nodes can never be adjacent, so every edge joins either two
reachable nodes or a reachable and an unreachable one. At day $118$ the split
is $16.2\%$ against $83.8\%$, and reachable nodes are the endpoints of $58.1\%$
of all edge endpoints. The network thus has a core and periphery structure, and
this appears directly in the \textit{degree assortativity}
coefficient~\cite{newman2002assortative}, which falls monotonically from
$-0.412$ at day $13$ to $-0.569$ at day $119$: the reachable core saturates
while the periphery keeps growing, so each reachable node absorbs more inbound
connections and the degree gap across edges widens. Despite this skew the
network remains extremely well connected. The giant component holds $99.72\%$
of nodes on average and the estimated diameter stays at $5$ throughout the
simulation. The fact that peer tables are dominated by unusable addresses
therefore do not translate into a fragile topology, because connection
formation is biased toward recently verified addresses. 

\subsubsection{The role of unreachable peers.}
To isolate what unreachable peers contribute we ran the same simulator with
every joining node reachable, targeting ten thousand nodes, for $704$ simulated
days. The runs share code and pipeline, so Table~\ref{tab:sim-comparison} is
age matched at day $118$ on identical instruments. The address level contrast
is stark. With every node reachable, a third of table entries had been
connected to successfully and the tried table stands at $29.6\%$ of capacity.
With $83\%$ of peers unreachable, both collapse, to $2.4\%$ and $7.5\%$,
because verification requires a successful connection that most addresses can
never provide. The same asymmetry governs what circulates, since $94.3\%$ of
cached addresses belong to nodes no connection can reach, against none in the
baseline. The topological contrast is sharper still
(Figure~\ref{fig:assortativity}). The all reachable network is mildly
assortative at $+0.055$ and stays there through day $704$. The subgraph of the
mixed network induced by its reachable nodes measures $+0.056$ at the same age,
indistinguishable from the baseline, while the full mixed graph measures
$-0.569$. The disassortativity is therefore not a property of the core under
load. It is produced entirely by the degree $8$ periphery attaching to it, and
since every outbound connection must terminate at a reachable node, the
periphery also forces the core toward the connection limit, raising the maximum
degree from $42$ to the cap of $125$.

\section{Conclusions}\label{sec:concl}
We investigated how the local peer table rules of Bitcoin-core shape the global
structure of the network, combining six months of crawl data from Bitcoin and
Dogecoin with a Bitcoin-core network-formation protocol simulator.  With our
empirical analysis we estimate network size, churn rate and address visibility
patterns in the data that is possible to collect with network crawlers.  Our
simulations, tuned according to the crawled data, shed light on three aspects
of the peer discovery process that are not observable by measurement: the
nature of the addresses that populate peer tables, the origin of the gap
between discovered addresses and active nodes observed, and the global topology
itself.

\begin{credits}
\subsubsection{\ackname}
This work was partly supported by the Research Council of Finland, Grant
363558, and by the University of Rome ``Tor Vergata'', RSA2024
cup:E83C25000860005.
\end{credits}

\bibliographystyle{splncs04}
\bibliography{psp}

\newpage\appendix
\begin{center}
\begin{Large}
\textbf{\LARGE Appendix}
\end{Large}
\end{center}

\section{Empirical analysis: supplementary material}
\label{app:empirical}

\subsection{Data collection and observables}
\label{app:data_collection}

We collect daily crawl data from the Bitcoin and Dogecoin peer-to-peer networks over two three-month periods (July--September 2025 and February--April 2026). For each network, a crawler connects to participating nodes and interacts with them using the standard peer-discovery protocol.

On each day, the crawler attempts to connect to previously known node addresses and records all nodes that complete the protocol handshake and return crawl data; we refer to these as \emph{crawled nodes}. For each such node, the crawler sends a \texttt{GETADDR} message, which requests a sample of the node's known peer addresses. In response, the node returns an \texttt{ADDR} message containing a random subset of entries from its internal peer table, with a default maximum of 1{,}000 addresses.

Importantly, peer tables represent a node's knowledge of other network participants rather than its current set of active connections. Consequently, repeated crawl responses from the same node provide multiple approximately independent samples from that node's local view of the network over time.

For each network and each day, we record:
\begin{itemize}
    \item the set of nodes that returned crawl data (crawled nodes);
    \item the peer-table samples contained in the \texttt{ADDR} responses;
    \item the union of all unique peer addresses observed across samples on that day (discovered peers); and
    \item the subset of those addresses that responded to an in-protocol \texttt{PING} message, which we use as a coarse indicator of availability.
\end{itemize}

Some nodes respond to protocol-level \texttt{PING} messages without returning \texttt{ADDR} data, so the set of \texttt{PING}-responsive nodes may strictly contain the set of crawled nodes. Throughout the analysis, we distinguish between these notions of participation where relevant.

Peer addresses are canonicalized and represented either as unique IP addresses or as \texttt{IP:port} pairs, depending on the analysis. Although IPv4 and IPv6 belong to distinct address families, peer tables frequently contain a mixture of both. We therefore include both address families in all empirical analyses, and merge addresses observed on different days to track node presence over time.

\subsection{Network size and session structure}
\label{app:session_length}

\begin{figure}[t]
\centering
\includegraphics[width=\linewidth]{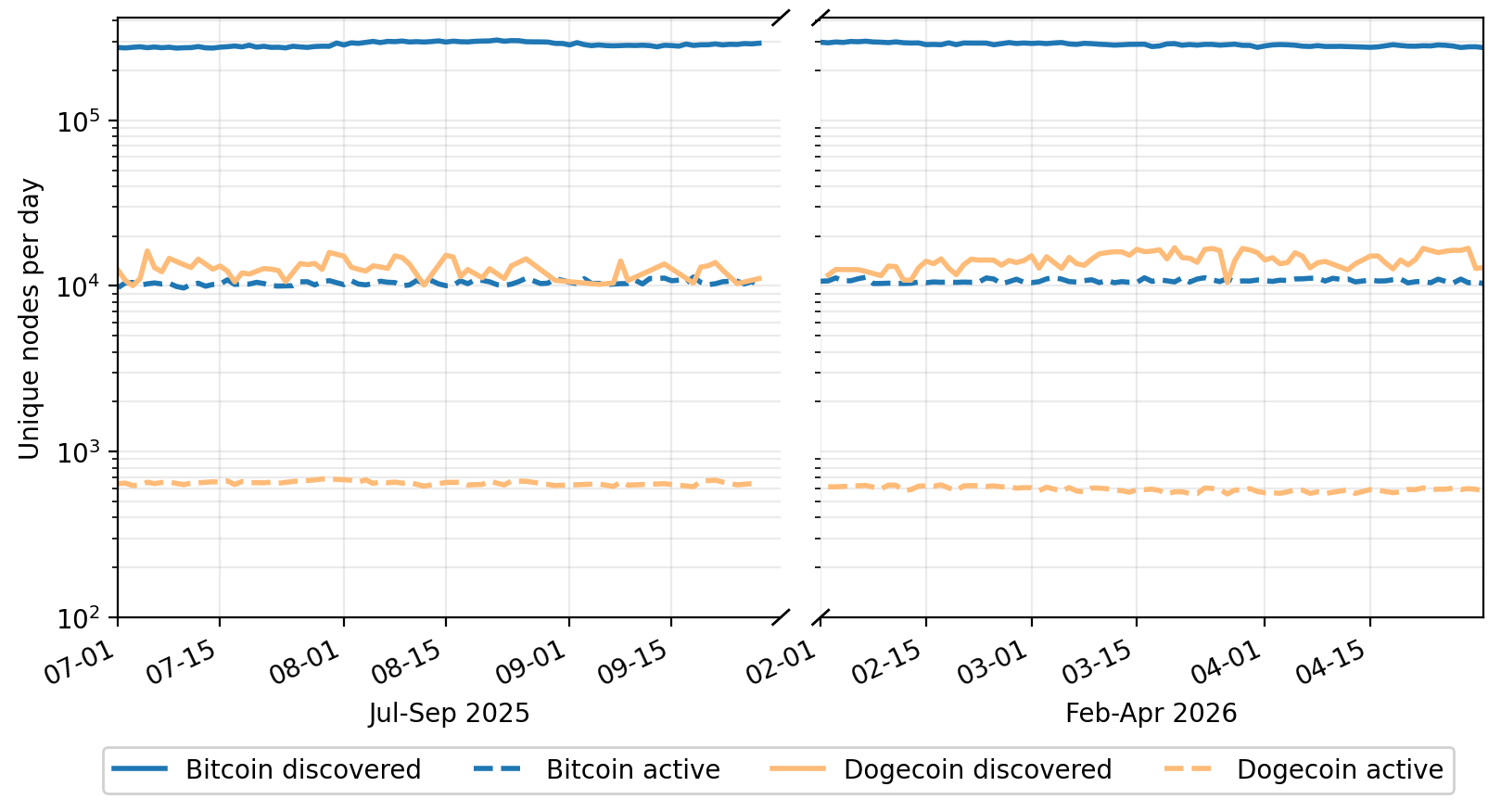}
\caption{Daily number of discovered and active nodes, identified by \texttt{IP:port}, for Bitcoin and Dogecoin. ``Discovered'' nodes are the union of peer-table entries observed on each day; ``active'' nodes are those responding to at least one in-protocol PING. The y-axis is logarithmic.}
\label{fig:daily-discovered-active}
\end{figure}

\begin{table}[h!]
\centering
\scriptsize
\setlength{\tabcolsep}{4pt}
\begin{tabular}{lllrrrrrrr}
\hline
Key & Period & Net. & Act./d & Disc./d & Act. & Disc. & IPv6$_A$ & IPv6$_D$ & Churn \\
\hline
\multirow{4}{*}{IP:port}
 & \multirow{2}{*}{Jul--Sep '25} & BTC  & 10{,}367 & 284{,}996 & 29{,}537 & 1{,}411{,}231 & 0.111 & 0.219 & 0.057 \\
 &                               & DOGE &    \,651 &  12{,}715 &  1{,}780 &    58{,}007   & 0.194 & 0.243 & 0.058 \\
 & \multirow{2}{*}{Feb--Apr '26} & BTC  & 10{,}696 & 286{,}578 & 30{,}903 & 1{,}031{,}468 & 0.160 & 0.199 & 0.052 \\
 &                               & DOGE &    \,590 &  14{,}338 &  1{,}710 &    46{,}765   & 0.186 & 0.269 & 0.053 \\
\hline
\multirow{4}{*}{IP}
 & \multirow{2}{*}{Jul--Sep '25} & BTC  & 10{,}297 & 275{,}344 & 29{,}112 & 1{,}042{,}699 & 0.112 & 0.228 & 0.057 \\
 &                               & DOGE &    \,618 &  12{,}607 &  1{,}700 &    57{,}381   & 0.197 & 0.244 & 0.058 \\
 & \multirow{2}{*}{Feb--Apr '26} & BTC  & 10{,}560 & 277{,}438 & 30{,}373 &    995{,}267  & 0.161 & 0.204 & 0.052 \\
 &                               & DOGE &    \,564 &  14{,}177 &  1{,}621 &    46{,}271   & 0.191 & 0.271 & 0.052 \\
\hline
\end{tabular}
\caption{Summary of network size and availability under two node-identity models: \emph{IP:port} and \emph{unique IP}, across two three-month measurement periods. Act./d and Disc./d denote median daily counts of active (PING-responsive) and discovered nodes; Act. and Disc. denote the corresponding totals (unique nodes observed at least once over the period). IPv6$_A$ and IPv6$_D$ are the median daily fractions of IPv6 addresses among active and discovered nodes, respectively. Churn is the mean daily drop rate computed with a one-day grace period (see Section~\ref{sec:empirical}); values are consistent across periods and close to $0.05$.}
\label{tab:network-summary}
\end{table}

Table~\ref{tab:network-summary} summarizes the overall scale of each network across both measurement periods. Bitcoin is substantially larger in both periods, with a median of over 10{,}000 active nodes per day and over one million unique addresses discovered per three-month period. Dogecoin is roughly an order of magnitude smaller, with approximately 590--650 active nodes per day. Key scale and churn metrics are consistent across both periods approximately six months apart, indicating that both networks are in a stable regime. Daily active and discovered node counts within each period are shown in Figure~\ref{fig:daily-discovered-active}.

\begin{figure}[b!]
\centering
\includegraphics[width=.9\linewidth]{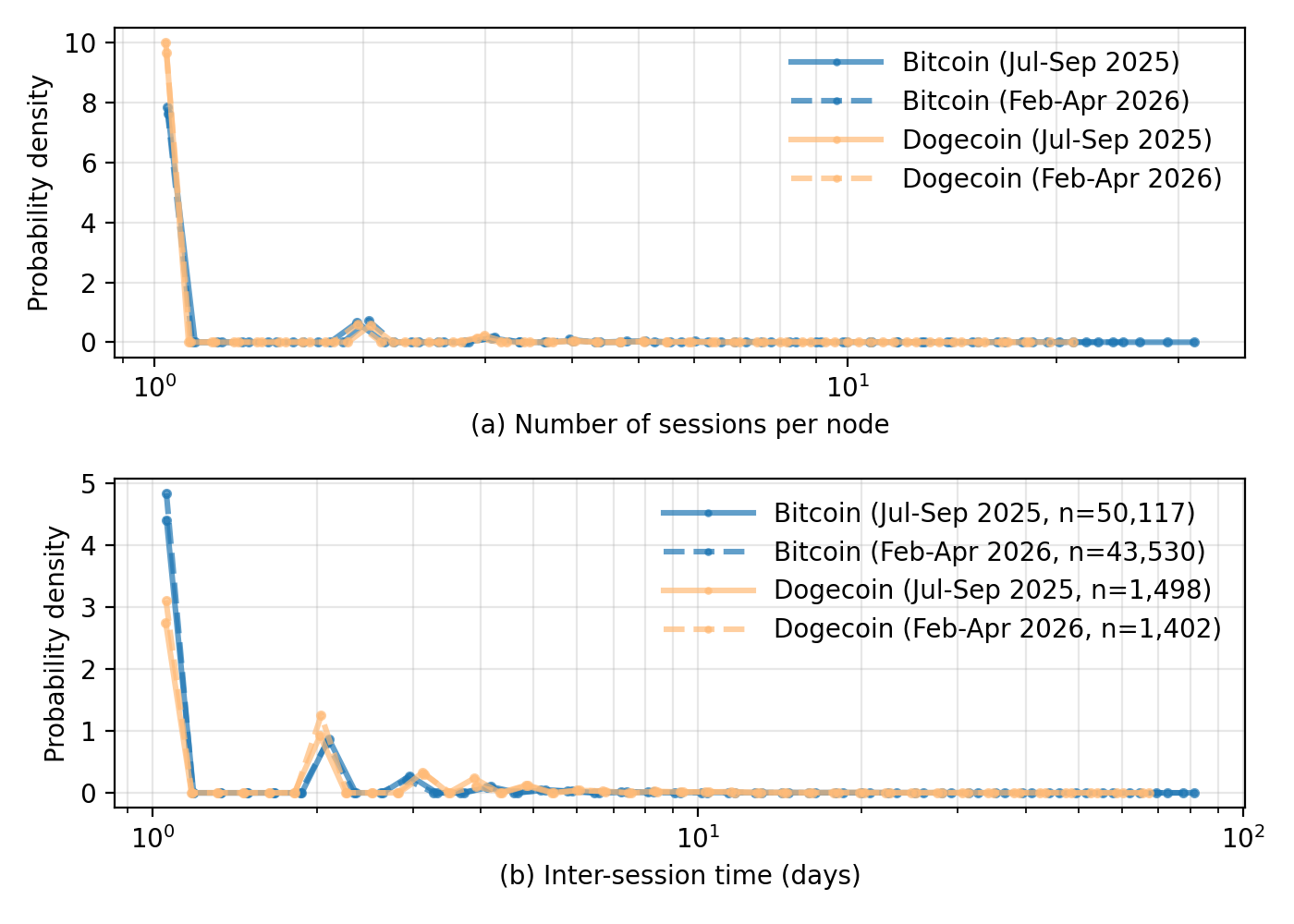}
\caption{Distribution of node sessions and inter-session gaps. Panel~(a): most nodes participate in a single continuous session. Panel~(b): among nodes with multiple sessions, inter-session gaps are strongly concentrated at one day, suggesting transient measurement artifacts rather than genuine node departures.}
\label{fig:session-dynamics}
\end{figure}

Figure~\ref{fig:session-dynamics} breaks down node participation into sessions. The majority of nodes appear in a single uninterrupted session, and among nodes with multiple sessions the gap between them is almost always exactly one day. This concentration at a one-day gap is the basis for the grace-period definition of churn used in Section~\ref{sec:empirical}: absences of a single day are treated as within-session rather than as genuine departures.

\subsection{Data cleaning and preprocessing}
\label{app:data_cleaning}

Our crawler indexes nodes by their IP address and port pair. To verify that responses from distinct ports on the same IP carry independent information, we compare the total number of crawl responses per IP with the number of unique responses (determined by hashing each response's peer set). Uniqueness rates are consistently high across both periods: 99.5\% of 664{,}904 Bitcoin responses in Jul--Sep 2025 and 99.2\% of 702{,}587 in Feb--Apr 2026; 98.1\% of 5{,}889 Dogecoin responses in Jul--Sep 2025 and 97.3\% of 6{,}303 in Feb--Apr 2026. This confirms that responses from different ports on the same IP are overwhelmingly distinct. We therefore retain the IP:port pair as the node identifier and restrict to purely unique responses per IP:port.

\begin{figure}[t!]
  \centering
  \includegraphics[width=.8\columnwidth]{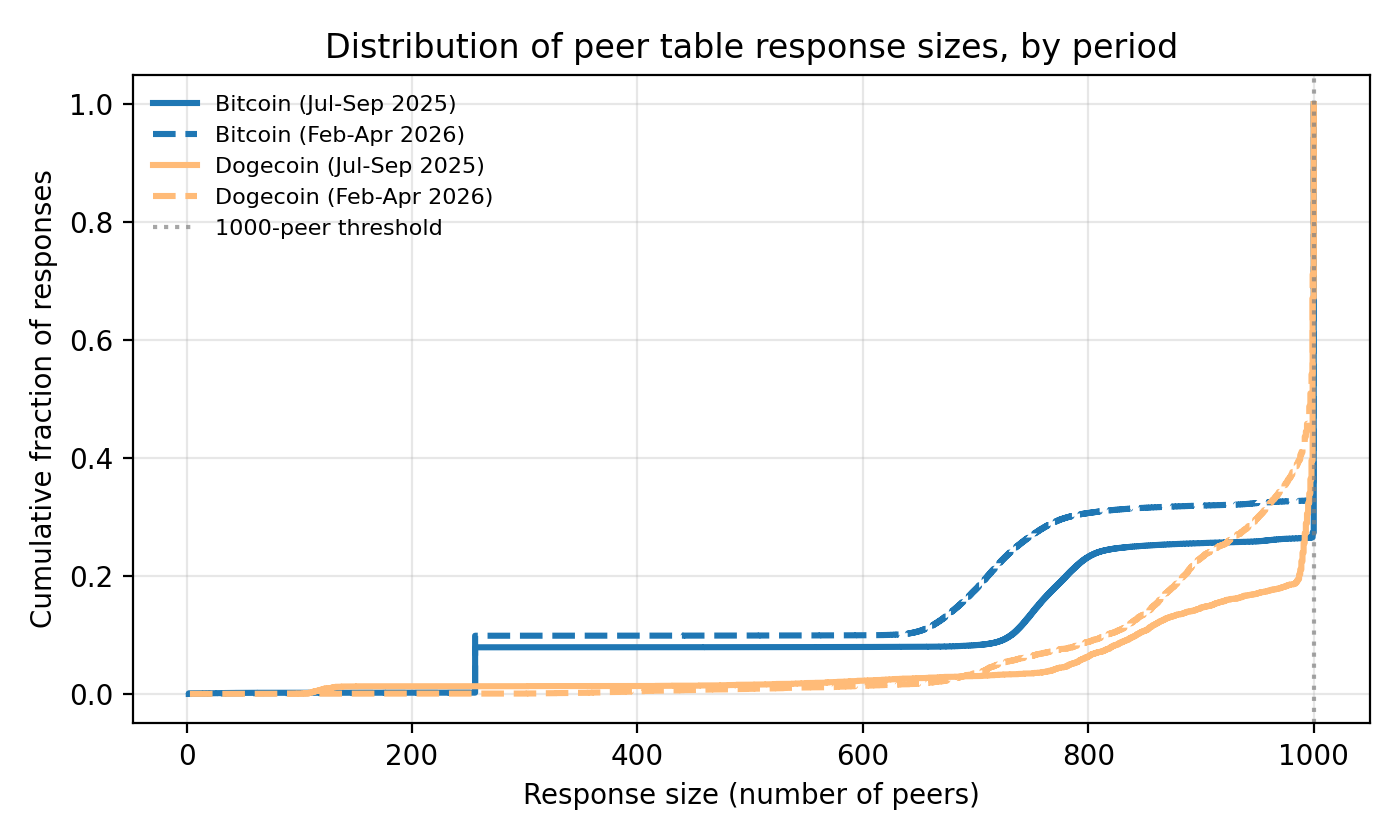}
  \caption{ECDF of crawl response sizes after deduplication, shown separately for each measurement period. Bitcoin responses concentrate near the protocol maximum of 1{,}000 addresses (approximately 65--70\% reach exactly 1{,}000 in both periods). Dogecoin responses are more variable (60--80\%), with the more recent period showing a trend toward smaller responses, likely reflecting a growing share of non-IPv4/IPv6 nodes.}
  \label{fig:response_size_dist}
\end{figure}

Figure~\ref{fig:response_size_dist} shows the distribution of \texttt{ADDR} response sizes after deduplication. Bitcoin responses are concentrated near the protocol maximum of 1{,}000 addresses in both periods (approximately 65--70\% reach exactly 1{,}000). Dogecoin responses are more variable across periods (60--80\%), with the more recent period showing a trend toward smaller responses. We restrict our analysis to responses containing at least 800 addresses, retaining 76.8\% of Bitcoin responses in Jul--Sep 2025 (508{,}039 of 661{,}293) and 69.3\% in Feb--Apr 2026 (483{,}416 of 697{,}230), and 93.7\% of Dogecoin responses in Jul--Sep 2025 (5{,}413 of 5{,}776) and 91.2\% in Feb--Apr 2026 (5{,}593 of 6{,}133). Smaller responses likely reflect nodes with Tor or other non-IPv4/IPv6 configurations, whose peer tables contain fewer addresses visible to our crawler (see Section~\ref{sec:background}); the majority of nodes run in standard IPv4/IPv6 configurations and return responses near the protocol maximum.

\section{Simulation: supplementary material}\label{sec:apx:simulation}

\subsection{The all reachable baseline}
\label{app:sim_baseline}

To separate the effects of unreachable peers from those of scale and maturation, we ran the simulator a second time with the reachability probability set to one, so that every joining node accepts inbound connections. All other mechanisms are identical, the arrival rate targets ten thousand nodes, and the run covers $704$ simulated days, stabilizing at approximately $10{,}079$ live nodes with a daily churn of $5.3\%$. Both runs are processed by the same extraction pipeline, so every quantity below is computed by the same code on both.

\begin{figure}[t]
    \centering
    \includegraphics[width=\linewidth]{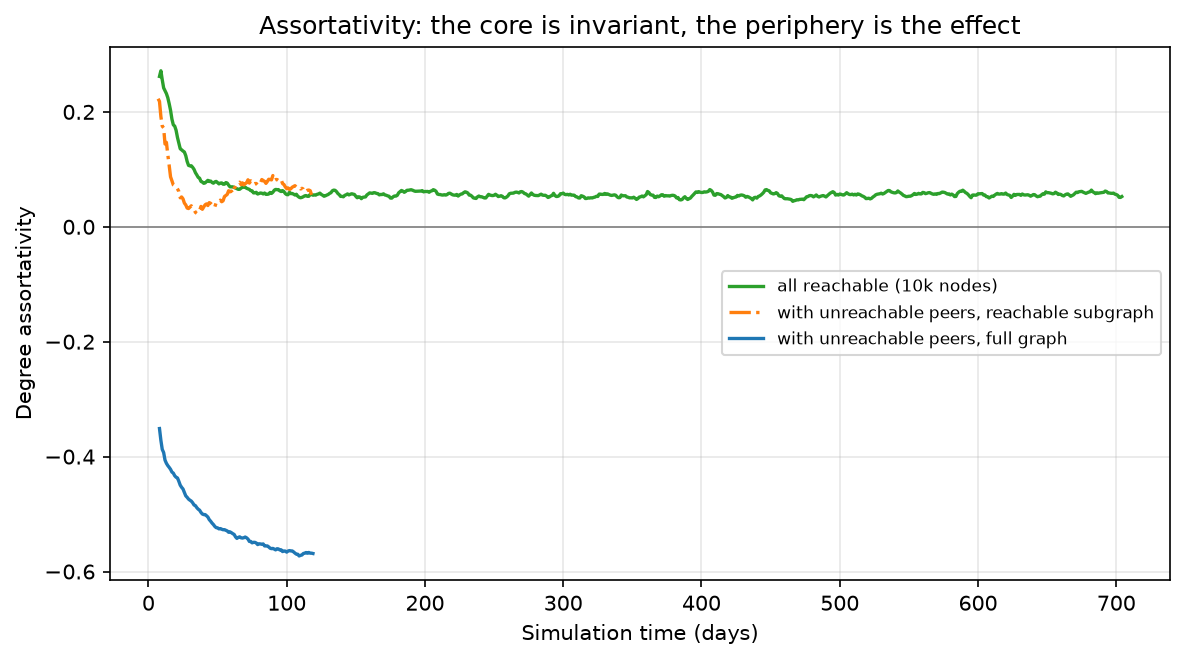}
    \caption{Degree assortativity over simulation time. The horizontal axis is the simulated day and the vertical axis Newman's assortativity coefficient~\cite{newman2002assortative}, ranging from $-1$ (disassortative) to $+1$ (assortative). The upper curve is the all reachable baseline over $704$ days, the lower curve the mixed network over its $120$ day window, and the dash dot curve the subgraph of the mixed network induced by its reachable nodes. The core tracks the baseline across the whole overlapping window while the full graph is driven to $-0.57$ by the degree $8$ periphery.}
    \label{fig:assortativity}
\end{figure}

Table~\ref{tab:sim-comparison} compares the two runs at day $118$, together with the subgraph of the mixed network induced by its reachable nodes. Three observations organise the comparison. First, the reachable core is invariant. The baseline network measures a degree assortativity of $+0.055$ at day $118$ and remains between $+0.05$ and $+0.06$ through day $704$, and the reachable induced subgraph of the mixed network measures $+0.056$ at the same age, while the full mixed graph measures $-0.569$ (Figure~\ref{fig:assortativity}). Second, the address tables invert. In the baseline the tried table, which holds only addresses verified by a successful connection, stands at $29.6\%$ of capacity against $18.8\%$ for the new table, and in the mixed run the ordering reverses to $7.5\%$ and $76.1\%$ because most circulating addresses can never be verified. Third, the baseline values at day $118$ persist essentially unchanged to day $704$, with the cached to live ratio moving from $2.20$ to $2.18$ and coverage from $45.4\%$ to $46.0\%$, which indicates that the day $118$ readings of both runs are equilibrium values rather than transients.

\begin{table}[t]
    \centering
    \scriptsize
    \setlength{\tabcolsep}{4pt}
    \begin{tabular}{lrrr}
        \hline
                                                     & all reachable & \multicolumn{2}{c}{with unreachable peers}                      \\
        Metric (day $118$)                           & $10$k target  & full graph                                 & reachable subgraph \\
        \hline
        live nodes                                   & $10{,}000$    & $33{,}655$                                 & $5{,}669$          \\
        unreachable share                            & $0\%$         & $83.2\%$                                   & $0\%$              \\
        mean degree                                  & $16.0$        & $16.0$                                     & $16.0$             \\
        median degree                                & $15$          & $8$                                        & $11$               \\
        maximum degree                               & $42$          & $125$                                      & $51$               \\
        giant component                              & $100\%$       & $100\%$                                    & $100\%$            \\
        estimated diameter                           & $5$           & $5$                                        & $5$                \\
        degree assortativity                         & $+0.055$      & $-0.569$                                   & $+0.056$           \\
        distinct cached addresses per live node      & $2.20$        & $8.60$                                     &                    \\
        distinct cached addresses per reachable node & $2.20$        & $2.91$                                     &                    \\
        cache coverage, end of day                   & $45.4\%$      & $11.6\%$                                   &                    \\
        peer table accuracy at departure             & $53.1\%$      & $17.2\%$                                   &                    \\
        table entries ever connected                 & $33.8\%$      & $2.4\%$                                    &                    \\
        new table entries                            & $12{,}299$    & $49{,}891$                                 &                    \\
        tried table entries (of $16{,}384$)          & $4{,}842$     & $1{,}222$                                  &                    \\
        \hline
    \end{tabular}
    \caption{The same simulator with and without unreachable peers, age matched at day $118$ and measured by the same extraction code. The last column restricts the mixed network to the subgraph induced by its reachable nodes. Address level quantities are properties of the whole population and are not repeated for the subgraph. The baseline values persist essentially unchanged to day $704$.}
    \label{tab:sim-comparison}
\end{table}
\begin{figure}[htb!]
    \centering
    \includegraphics[width=0.8\columnwidth]{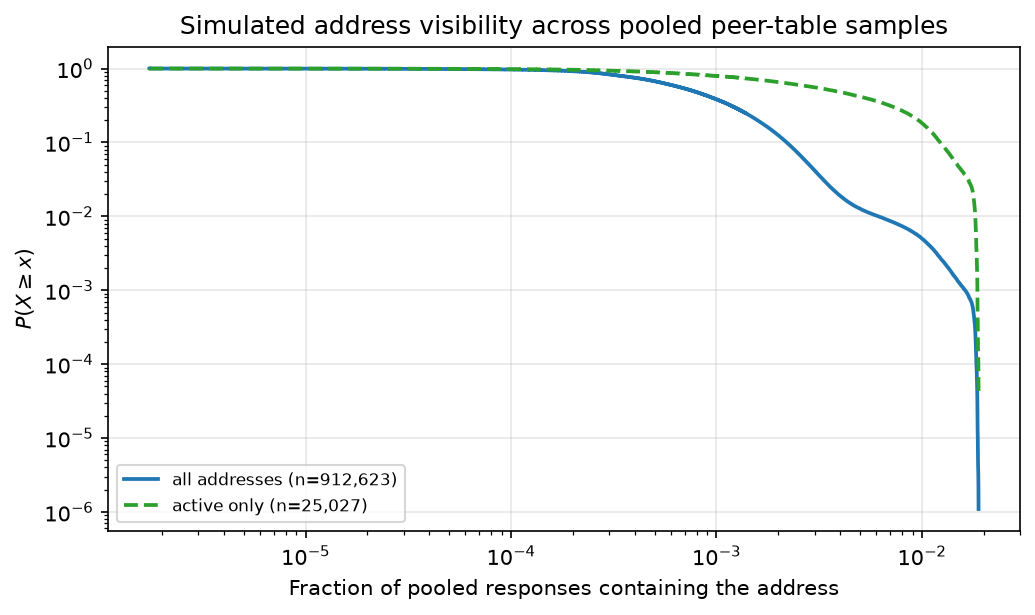}
    \caption{CCDF of simulated address visibility under definition~A, pooled
    over the $575{,}609$ full size responses of days $28$ to $120$. The
    horizontal axis is the fraction of pooled responses containing the address
    and the vertical axis the probability of reaching at least that fraction.
    Active addresses are systematically more visible than the full population,
    matching the qualitative structure of Figure~\ref{fig:address-frequency}.}
    \label{fig:crawl_visibility}
\end{figure}

\subsection{Where the inbound load goes}
\label{app:sim_saturation}

The degree contrast in Table~\ref{tab:sim-comparison} follows from conservation of connections. Every node opens $8$ outbound connections and unreachable nodes accept none, so all $8 \times 27{,}986 \approx 224{,}000$ outbound connections of the periphery terminate on the $5{,}669$ reachable nodes, an average of about $39$ inbound connections each, on top of their own $8$ outbound and roughly $8$ inbound from other reachable nodes. This accounts for the observed mean reachable degree of $51.7$. The load is not spread evenly, since long lived nodes accumulate inbound connections over their session, which produces the second mode at the connection limit of $125$ in Figure~\ref{fig:degree_distribution}. In the baseline the same inbound volume is spread over the whole population, the mean inbound is $8$, the daily maximum degree stays near $42$, and no node approaches the limit.

\subsection{Supplementary figures}
\label{app:sim_figures}
\begin{figure}[t]
    \centering
    \includegraphics[width=\columnwidth]{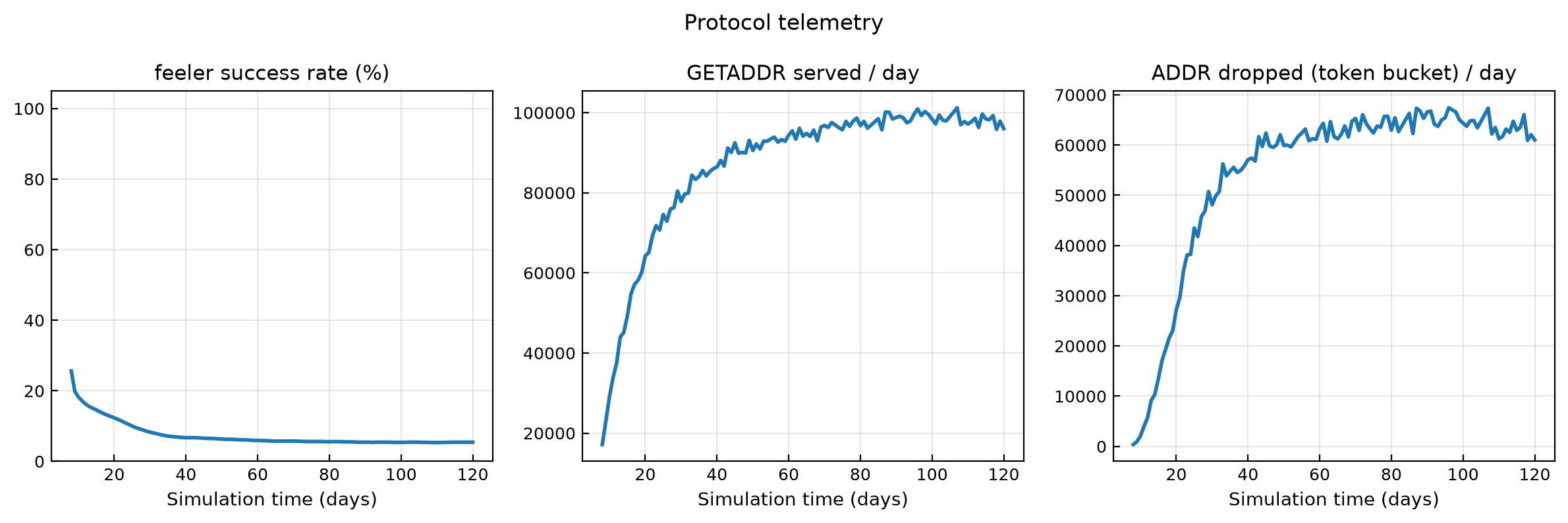}
    \caption{Protocol telemetry over days $7$ to $119$ of the mixed run. Feeler success rate (left), \texttt{GETADDR} responses served per day (center), and unsolicited addresses discarded per day by the token bucket (right).}
    \label{fig:telemetry}
\end{figure}

\begin{figure}[t]
    \centering
    \includegraphics[width=\columnwidth]{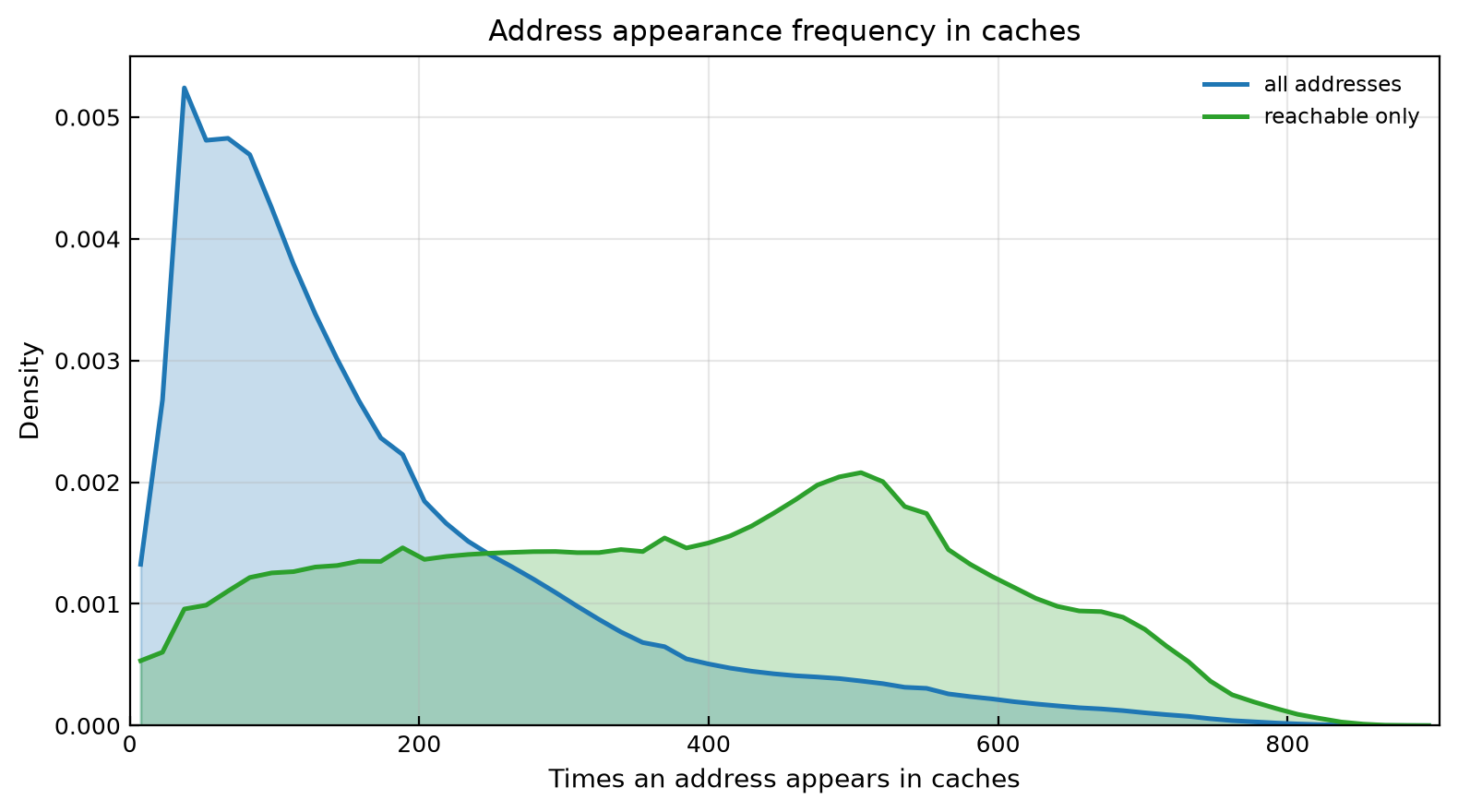}
    \caption{Distribution of the number of caches an address appears in on a single day, averaged over the $104$ stabilized days of the mixed run. The horizontal axis is the daily appearance count and the vertical axis the density over addresses. One curve covers all cached addresses and the other only those of reachable nodes, which are carried in about $2.2$ times as many caches.}
    \label{fig:ip_frequency}
\end{figure}

\begin{figure}[t]
    \centering
    \includegraphics[width=0.8\columnwidth]{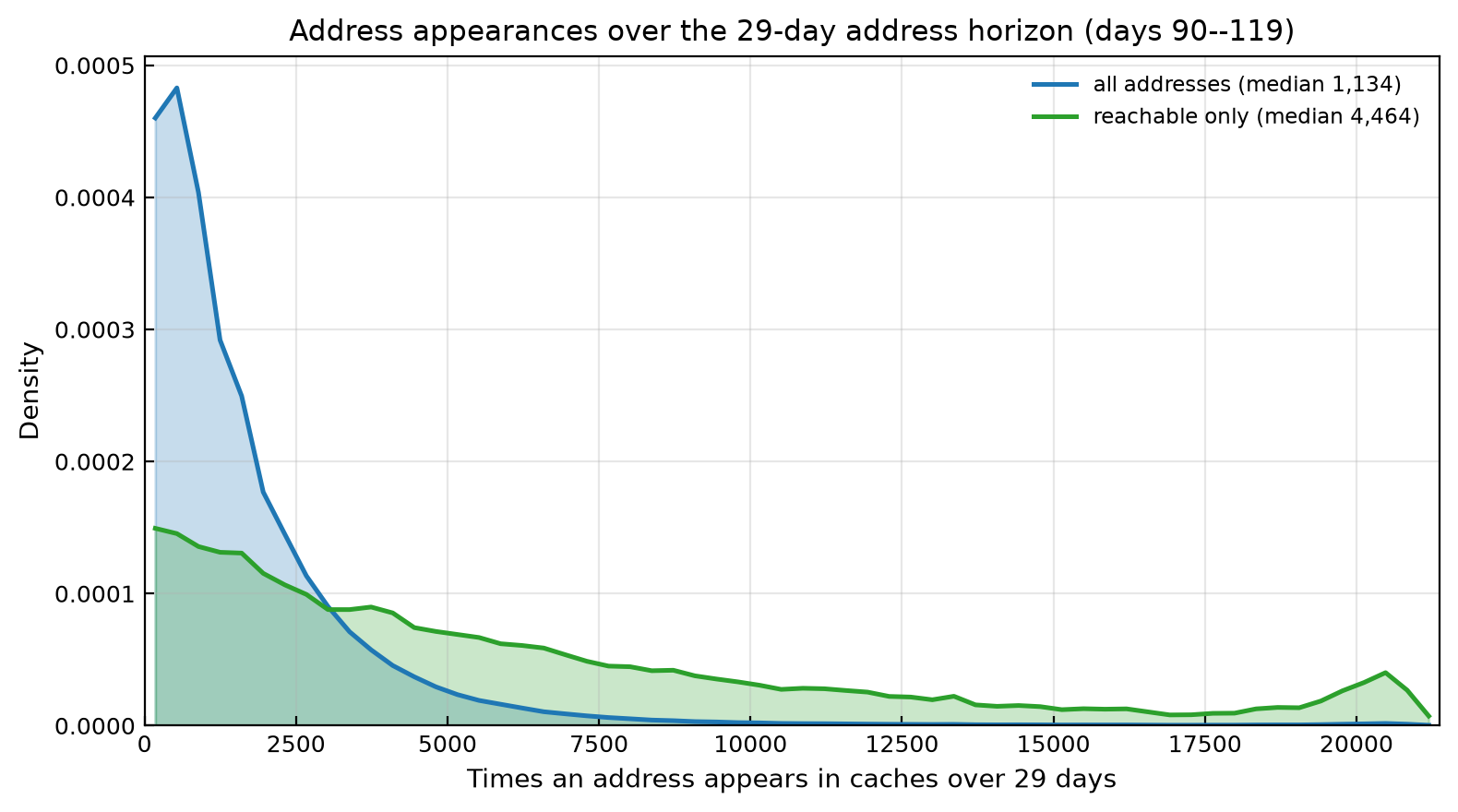}
    \caption{Number of caches holding each address, totalled over the trailing $29$ days of the window. The horizontal axis is the total appearance count and the vertical axis the density over addresses. One curve covers all cached addresses and the other only those belonging to reachable nodes, whose median is $3.9$ times higher. The rise at the right edge is the saturation ceiling, addresses present in essentially every cache throughout, all of them reachable.}
    \label{fig:ip_frequency_horizon}
\end{figure}

Figure~\ref{fig:telemetry} shows the protocol
telemetry over the window, and Figures~\ref{fig:ip_frequency}
and~\ref{fig:ip_frequency_horizon} the cache appearance distribution split by
class, per day and over the address horizon, all supporting the discussion of
Section~\ref{sec:simulations}.

Aggregating the daily counts of Figure~\ref{fig:ip_frequency} over the trailing thirty days rather than per day sharpens the separation (Figure~\ref{fig:ip_frequency_horizon}).Over that window a reachable address is held in $3.9$ times as many caches as the median address, $4{,}464$ against $1{,}134$, where the daily view gave $2.2$. A persistent advantage compounds across a month, whereas a single day is dominated by which caches happened to rebuild. The concentration is stronger than the means suggest, since reachable addresses are $4.1\%$ of the population and carry $14.7\%$ of all appearances, yet they make up $89.4\%$ of the most visible percentile. At the right edge sits a small population of $1{,}256$ addresses, $0.22\%$ of the total, held in almost every cache for the entire window, and every one of them is reachable. These are the long lived nodes that also produce the second mode at the connection limit in Figure~\ref{fig:degree_distribution} of Section~\ref{sec:simulations}.

\end{document}